\documentclass{ifacconf}
\usepackage{graphicx}      % include this line if your document contains figures
\usepackage{svg}
\usepackage{natbib}        % required for bibliography
\usepackage{array}
\usepackage{booktabs}
\usepackage{tcolorbox}
\usepackage{amssymb}
\usepackage{balance}
\usepackage{amsmath,amssymb,amsfonts}
\usepackage{algorithmic}
\usepackage{textcomp}
\usepackage[dvipsnames]{xcolor}
\usepackage{csquotes}
\def\BibTeX{{\rm B\kern-.05em{\sc i\kern-.025em b}\kern-.08em
    T\kern-.1667em\lower.7ex\hbox{E}\kern-.125emX}}
\usepackage{mathtools}

\usepackage{circuitikz}

\newcommand{\xb}{\mathbf{x}}
\newcommand{\xbd}{\dot{\mathbf{x}}}
\newcommand{\xbt}{\mathbf{x}(t)}
\newcommand{\xbh}{\hat{\mathbf{x}}}
\newcommand{\ub}{\mathbf{u}}
\newcommand{\ubt}{\mathbf{u}(t)}
\newcommand{\yb}{\mathbf{y}}
\newcommand{\ybt}{\mathbf{y}(t)}
\newcommand{\ybh}{\hat{\mathbf{y}}}
\newcommand{\thb}{\mathbf{\theta}}

\newcommand{\jb}{\mathbf{J}}
\newcommand{\jbx}{\mathbf{J}(\xb)}
\newcommand{\jbth}{\mathbf{J}_{\theta}}
\newcommand{\rb}{\mathbf{R}}
\newcommand{\rbx}{\mathbf{R}(\xb)}
\newcommand{\rbth}{\mathbf{R}_{\theta}}
\newcommand{\gb}{\mathbf{G}}
\newcommand{\gbx}{\mathbf{G}(\xb)}
\newcommand{\gbth}{\mathbf{G}_{\theta}}
\newcommand{\pb}{\mathbf{P}}
\newcommand{\pbx}{\mathbf{P}(\xb)}
\newcommand{\pbth}{\mathbf{P}_{\theta}}
\newcommand{\hx}{H(\xb)}
\newcommand{\hth}{H_{\theta}}
\newcommand{\qtilde}{\mathbf{\tilde{Q}}}
\begin{document}
\begin{frontmatter}

\title{Improved Initialization for Port-Hamiltonian Neural Network Models\thanksref{footnoteinfo}} 

\thanks[footnoteinfo]{This work is funded by the European Union (Horizon Europe, ERC, COMPLETE, 101075836). Views and opinions expressed are however those of the author(s) only and do not necessarily reflect those of the European Union or the European Research Council Executive Agency. Neither the European Union nor the granting authority can be held responsible for them.}

\author[First]{G.J.E. van Otterdijk} 
\author[First]{S. Weiland}
\author[First]{M. Schoukens}

\address[First]{Control Systems Group, Eindhoven University of Technology, Eindhoven, the Netherlands.}

\begin{abstract}
Port-Hamiltonian neural networks have shown promising results in the identification of nonlinear dynamics of complex systems, as their combination of physical principles with data-driven learning allows for accurate modelling. However, due to the non-convex optimization problem inherent in learning the correct network parameters, the training procedure is prone to converging to local minima, potentially leading to poor performance. In order to avoid this issue, this paper proposes an improved initialization for port-Hamiltonian neural networks. The core idea is to first estimate a linear port-Hamiltonian system to be used as an initialization for the network, after which the neural network adapts to the system nonlinearities, reducing the training times and improving convergence. The effectiveness of this method is tested on a chained mass-spring-damper setup for varying noise levels and compared to the original approach.
\end{abstract}


\end{frontmatter}
%===============================================================================
\section{Introduction}
% Describe the general problem of SI, why do we want models and why is it difficult to obtain them
Accurate models of dynamical systems are essential in understanding and controlling complex systems. Acquiring these models is not a trivial task however, especially in the presence of nonlinear dynamics and measurement noise \citep{SchoukensLjung2019}. It is thus vital to select the right modelling approach, as that choice significantly influences the quality of the obtained model.

For instance, while solely data-based methods require little to no expert knowledge, they are often unable to provide physical guarantees. On the other hand, methods based purely on physical knowledge might only be relevant for a small set of systems, limiting their range of application.
It is therefore no surprise that
%recent methods have attempted
%there has been a trend recently 
%research has been focused on
efforts have been made to
combine physical knowledge with the available data to achieve the best of both worlds \citep{GREY_bohlin, PINN_survey}.

% Par 2 - PHNNs are a nice promising SI method
%Among these so-called grey-box modelling techniques, physics-informed neural networks have seen extensive research recently \citep{PINN_survey, PINN_stiasny, PINN_jia}, and have led to state-of-the-art results in various application fields \citep{PIN_SOTA_gotte}.

%\textcolor{red}{Find additional citations/reviews.} 
A particular approach worth mentioning is the output-error port-Hamiltonian neural network (PHNN) \citep{PHNN_moradi}. Inspired by earlier work on Hamiltonian neural networks \citep{HNN_greydanus}, it is based on the underlying port-Hamiltonian (PH) theoretical framework \citep{PHS_vdschaft} and offers several benefits; it allows for multi-physics modelling, ensures that the identified models are passive, and also yields a modular description of interconnected systems. Previous work has shown that this modularity can be exploited to combine known and unknown dynamics within the same system \citep{PHNN_otterdijk}.
%Among these grey-box modelling techniques, physics-informed neural networks have seen great performance and are applicable to a diverse range of systems \textcolor{red}{CITE}. In particular, the development of the Port-Hamiltonian Neural Network (PHNN) \textcolor{red}{CITE} offers the modelling for a wide range of physical systems and has shown promising results thus far. It is based around the Port-Hamiltonian framework, which \textcolor{red}{Explain core ideas of PH modelling}

% Par 3 - PHNNs suffer from some drawbacks --> could they be solved through initialization?
At the same time, there remain concerns about the computational feasibility of these PHNNs.
In fact, due to the non-convex optimization problem inherent in learning the model parameters it is found that the estimated model does not reliably converge to the noise floor as expected. 
%In fact, it is found that the non-convex optimization problem of learning the model parameters prevents the estimated model from reliably converging to the noise floor.
% Par 4 - Indicate the existing research into initialization and linear PH methods
A potential area for improvement lies in the selection of initial parameters for the network. The literature shows that the choice of initialization can have a significant impact on the performance of the algorithm \citep{INIT_thesis, INIT_alt, INIT_marconato}. Among more recent results, the use of a linear estimate of the system seems promising in both the training convergence as well as the final accuracy of the obtained model \citep{INIT_schoukens1, INIT_schoukens2, INIT_floren}.

However, taking a similar approach for the PHNN requires a good linear port-Hamiltonian estimate, which is not a straightforward task. Existing work has shown how to calibrate a linear PH estimate based on data \citep{LINPH_claudia1, LINPH_claudia2}, but these methods do not allow for the identification of a linear PH system from scratch. As an alternative to the direct identification of linear PH models, a minimal positive real or passive representation of the system can be identified instead, since it is known that there exists a PH equivalent for such a system \citep{PRPH_cherifi}. Several methods exist to enforce the passivity of a linear state-space system \citep{PAreview_mahanta}, resulting in a similar variety of metrics for the closest passive system. An interesting approach is given in \citep{PR_gillis}, which considers the minimal distance between the system matrices of the PH system and the initial estimate. However, none of the examined approaches sufficiently match the input-output behaviour of the linear PH system and the linear estimate.
% Part of the existing method is selecting an initialization of the network. In the literature it can be found that the choice of initialization can have a significant impact on the performance of the algorithm \citep{LI_xavierinit, LI_erhan2009b, LI_erhan2010}. Among the results focusing on system identification, using a linear estimate of the system seems promising in both the training convergence as well as the final accuracy of the obtained model \citep{LI_maarten_LFRinit, LI_maarten_ssinit}. Therefore, this section details the process of finding and utilizing a linear PH estimate for the PHNN.

% Par 5 - Clearly state the gap in research + our contributions
To overcome these limitations, this paper proposes an alternative method to obtain linear PH estimates, which in turn can be used to enhance the reliability and computational feasibility of PHNNs. The main contributions of this work can be summarised as follows:
\begin{itemize}
    \item An initialization strategy for non-linear PHNNs
    \item Improved convergence and a better model fit for non-linear PHNNs
\end{itemize}

%GAP IN RESEARCH: Consistent good convergence of PHNNs for a wider variety of systems / experimental setup? / Feasibility of larger systems

% Par 5 - this paper is structured as follows
With regard to the rest of the paper, Section \ref{sec:preliminaries} introduces the nonlinear system identification problem, alongside important aspects of port-Hamiltonian systems theory and the underlying structure of the original PHNN. Afterwards, Section \ref{sec:linear initialization} details the process of obtaining a linear port-Hamiltonian estimate and its incorporation into the existing structure. To validate the performance of the initialization scheme, Section \ref{sec:results} studies two numerical examples and compares the proposed method to the original. Finally, Section \ref{sec:conclusion} provides a brief conclusion.

\section{Preliminaries}
\label{sec:preliminaries}
% Purpose of this section
Before going into detail on the proposed changes to the PHNN, the existing structure must be clearly detailed. This is done in two parts; firstly defining the identification problem and secondly recalling the existing PHNN structure \citep{PHNN_moradi}, based on the SUBNET identification approach \citep{MS_subnet}.

% Introducing the System identification task
\subsection{Identification Problem}
Consider a general continuous time system described by 
\begin{subequations}
    \begin{equation}
        \xbd(t) = f(\xbt, \ubt),
        \label{eq:f_function}
    \end{equation}
    \begin{equation}
        \ybt = h(\xbt),
        \label{eq:h_function}
    \end{equation}
\end{subequations}
where $\ubt \in \mathbb{R}^{n_{u}}$, $\xbt \in \mathbb{R}^{n_{x}}$ and $\ybt \in \mathbb{R}^{n_{y}}$ are the input, state and output vectors respectively, with $n_{u}$, $n_{x}$ and $n_{y}$ being the corresponding number of inputs, states and outputs. Data can be sampled from this system with a sampling time of $T_{s}$ according to $\mathbf{u}_{k} = \mathbf{u}(kT_{s})$ and $\mathbf{y}_{k} = \mathbf{y}(kT_{s}) + \varepsilon_{k}$, with $\varepsilon_{k}$ representing a zero-mean, finite-variance additive output measurement noise. This leads to a dataset of $N$ input-output samples,
\begin{equation}
    \mathcal{D}_N=\{(\mathbf{y}_{k}, \mathbf{u}_{k}\}_{k=0}^{N-1}.
    \label{eq:dataset}
\end{equation}
The objective for the system identification task is to find the model that best describes the dynamics of the underlying system given this set of datapoints.
% ZOH?

% Introducing the notation for Port-Hamiltonian systems
\subsection{Port-Hamiltonian Systems}
\label{sec:PHS}
As mentioned before, the port-Hamiltonian framework offers several advantages in the areas of multi-physics modelling, modularity and passivity. It has a rich theoretical background \citep{PHS_vdschaft}, but for most practical purposes a general input-state-output PH system without feedthrough can be described by
\begin{subequations}
    \begin{equation}
        \begin{aligned}
            \xbd(t) =& \bigl[ \jb(\xbt)-\rb(\xbt)\bigr]\nabla H(\xbt) \\
            &+ \bigl[\gb(\xbt) - \pb(\xbt)\bigr]\ubt,
        \end{aligned}
        \vspace{-1mm} % Dont leave an open line between the two equations
    \end{equation}
    \begin{equation}
        \ybt = \bigl[\gb(\xbt) + \pb(\xbt)\bigr]^{\top}\nabla H(\xbt),
    \end{equation}
    \label{eq:PHS}
\end{subequations}
where $\ubt \in \mathbb{R}^{n_{p}}$, $\xbt \in \mathbb{R}^{n_{x}}$ and $\ybt \in \mathbb{R}^{n_{p}}$ are again the input, state and output vectors respectively, with $n_{x}$ being the number of states and $n_{p}$ the number of ports. For the sake of notational clarity, the time dependence of these vectors will be left out from here on out.
Furthermore, $\jbx \in \mathbb{R}^{n_{x} \times n_{x}}$ is the skew-symmetric structure matrix, $\rbx \in \mathbb{R}^{n_{x} \times n_{x}}$ is the symmetric positive semi-definite dissipation matrix, $\hx \in \mathbb{R}$ is a scalar function representing the total energy of the system, while ${\gbx, \pbx} \in \mathbb{R}^{n_{x} \times n_{p}}$ are the external matrices.

% Introducing the Port-Hamiltonian Neural Network
\subsection{Port-Hamiltonian Neural Networks}
\label{sec:PHNN}
In order to identify a model for the continuous time system as shown in \eqref{eq:PHS}, from the given dataset \eqref{eq:dataset}, an output-error (OE) model is considered. This is done by following the SUBNET formulation, consisting of three main components. First, an initial state is estimated by an encoder, based on previous input-output data. Secondly, that state is propagated forward through time by numerically integrating the state-evolution equation. Finally, from this trajectory the outputs can be calculated using the output equation. By optimizing the parametrizations of these components based on a truncated simulation loss, this setup enables the identification of a continuous PH model, even in the absence of any state information in the dataset.

To match the structure provided in the PH framework, both the state evolution and the output equations can be represented by the following parametrised functions
\begin{subequations}
    \begin{equation}
    \label{eq:state_evol_param}
            \hat{\xbd} = \bigl[ \jb_{\theta}(\xbh)-\rb_{\theta}(\xbh)\bigr]\nabla H_{\theta}(\xbh) + \bigl[\gb_{\theta}(\xbh) - \pb_{\theta}(\xbh)\bigr]\ub
    \end{equation}
    \begin{equation}
    \label{eq:output_eq_param}
        \ybh = \bigl[\gb_{\theta}(\xbh) + \pb_{\theta}(\xbh)\bigr]^{\top}\nabla H_{\theta}(\xbh),
    \end{equation}
\end{subequations}
where the matrices $\jbth, \rbth, \gbth, \pbth$ and $\hth$ are constructed as \emph{multi-layer-perceptrons} (MLPs), all depending on the estimated state $\xbh$.

To ensure that the structural requirements introduced in Section \ref{sec:PHS} are met, the following transformations are performed on the output of the MLPs. $\jbth(\xb)$ is defined as the difference between a MLP output matrix $\mathbf{A}_{\theta}(\xb)$ and its transpose, i.e. $\jbth(\xb) = \mathbf{A}_{\theta}(\xb) - \mathbf{A}^{\top}_{\theta}(\xb)$, always resulting in a skew-symmetric parametrisation of $\jbth(\xb)$. In a similar fashion, $\rbth(\xb)$ is defined as the product of a MLP output matrix $\mathbf{B}_{\theta}(\xb)$ and its transpose, i.e. $\rbth(\xb) = \mathbf{B}_{\theta}(\xb)\mathbf{B}^{\top}_{\theta}(\xb)$, always resulting in a symmetric positive semi-definite parametrisation of $\rbth(\xb)$. Furthermore, to ensure that $H_{\theta}(\xb)$ is bounded from below, an exponential-linear unit (ELU) is applied to the output of the MLP responsible for its parametrisation. Finally, since $\gbth(\xb)$ and $\pbth(\xb)$ do not have requirements on their structure, they are directly represented by separate MLPs.

To learn the right parameters for these MLPs, the truncated simulation loss function from SUBNET is minimized:
\begin{subequations}
\label{eq:SUBNET}
\begin{align} 
    V_{\mathcal{D}_N}(\theta,\eta) &= \frac{1}{C} \sum_{t=n+1}^{N-T+1}\sum_{k=0}^{T-1} \left\|\yb_{t+k}- \ybh_{t+k|t}\right\|_2^2, \label{subeq:loss_theta}\\  
    \intertext{subject to:} 
    \xbh_{t|t} &= \psi_\eta (\yb_{t-n_{a}}^{t-1}, \ub_{t-n_{b}}^{t-1}), \label{subeq:x_theta}\\
    \xbh_{t+k+1|t} &= \text{ODE-solver} [f_{\theta}(\xbh_{t+k|t}, \ub_{t+k})], \label{subeq:x_k+1_theta}\\
    \ybh_{t+k|t} &= h_{\theta}(\xbh_{t+k|t}),    \label{subeq:y_theta}
\end{align}
\end{subequations}
where the pipe ($\mid$) notation is introduced to distinguish between subsections as (current index$\mid$start index), and $C = (N-T+1)T$. Additionally, the initial state of these subsections is estimated by the encoder $\psi_{\eta}$ based on past $n_{a}$ past input and $n_{b}$ past output samples. i.e.: $\ub_{t-n_{b}}^{t-1} = \begin{bmatrix} \ub_{t-n_{b}}^\top,  \cdots,  \ub_{t-1}^\top \end{bmatrix} ^\top$; with $\yb_{t-n_{b}}^{t-1}$ defined in a similar way. Because there is no access to state information in the available data, the encoder, state evolution and decoder must be trained at the same time.

\section{Linear PH Initialization}
\label{sec:linear initialization}
% Highlight that a good initialization can make a significant difference
% Detail the approach we take to find the closest linear PH system
% Go into normalisation / ensuring positive semi-definiteness
% Detail how a linear estimate can be combined with a nonlinear NN

% Refer to Claudia Totzeck's paper for normalization
This section discusses the initialization approach to the OE-PHNN, by first detailing the process of finding a linear PH estimate and next describing how this linear estimate is incorporated in the initialization.

\subsection{Obtaining a linear PH estimate}
% How do we find a suitable linear PH?
Two options are considered to determine a linear PH estimate. The first option makes use of the \emph{Best Linear Approximation} (BLA), which is a linear model estimate that achieves the best least squares fit on the output of the nonlinear system \citep{LI_BLA}. The second option directly identifies a linear PH estimate by restricting a PHNN to the linear case.

\textbf{Indirect BLA estimate:}
For the first option, the BLA model estimate serves as the starting point, yielding the state space matrices, $(\mathbf{A}, \mathbf{B}, \mathbf{C}, \mathbf{D})$, describing the linear system:
\begin{subequations}
    \begin{equation}
        \hat{\xbd} = \mathbf{A}\xb + \mathbf{B}\ub,
    \end{equation}
    \begin{equation}
        \ybh = \mathbf{C}\xb + \mathbf{D}\ub. 
    \end{equation}
\end{subequations}
% We want to find an equivalent PH system --> So we solve the KYP for Q.
% However, if the BLA does not deliver a passive linear estimate, the KYP cannot be solved.
% Instead, Q is found to be slightly negative? --> So we apply a projection onto a pos semi def matrix to guarantee a linear PH system.
% With this sym pos semi def Q, we can derive the other matrices
In order to convert this regular state space description to an equivalent PH system, the \emph{Kalman-Yakubovich-Popov} (KYP) inequality, which is nicely summarized in \citep{LI_KYP}, needs to be solved for the symmetric positive semi-definite matrix $\qtilde=\qtilde^{\top} \succeq 0$:
\begin{equation}
    \begin{bmatrix}
        -\mathbf{A}^{\top}\qtilde-\qtilde^{\top}\mathbf{A} &\hspace{2mm} \mathbf{C}^{\top}-\qtilde^{\top}\mathbf{B} \\
        \mathbf{C}-\mathbf{B}^{\top}\qtilde &\hspace{2mm} \mathbf{D}+\mathbf{D}^{\top}
    \end{bmatrix}
    \geq 0,% \hspace{10mm}
%    \mathbf{Q} = \mathbf{Q}^{\top},
\end{equation}
where the matrix $\qtilde$ represents the Hamiltonian function, which is found to be quadratic for linear PH systems.

However, the KYP inequality is only solvable for passive systems. Since the BLA does not guarantee such a passive estimate, there is also no guarantee that a positive semi-definite Hamiltonian, $\qtilde$, can be found, even when the nonlinear data-generating system is passive. In practice, it is observed that the BLA is more likely to yield non-passive estimates for highly nonlinear systems, or in cases of significant measurement noise. To circumvent this issue, the nearest symmetric positive semi-definite matrix $\mathbf{Q}$, to the original $\mathbf{\tilde{Q}}$, can be found by following the steps in \citep{LI_higham}.
% There is an issue here however. It is not guaranteed that the BLA matrices are passive. In that case, the LMI is solved up to a tolerance Q >= -tI. This means that the converted representation is not PH. Therefore, if the original Q < 0 (i.e. not pos semidef), we take its projection on the PSD cone instead: see paper ??

With the Hamiltonian matrix $\mathbf{Q}$ obtained, the structure, dissipation and external matrices are constructed as follows \citep{PRPH_cherifi}:
\begin{align*}
    \mathbf{J}&=\frac{1}{2}(\mathbf{A}\mathbf{Q}^{-1}-\mathbf{Q}^{-\top}\mathbf{A}^{\top}), & \mathbf{R}&=\frac{1}{2}(\mathbf{A}\mathbf{Q}^{-1}+\mathbf{Q}^{-\top}\mathbf{A}^{\top}),\\
    \mathbf{G}&=\frac{1}{2}(\mathbf{Q}^{-\top}\mathbf{C}^{\top}+\mathbf{B}), & \mathbf{P}&=\frac{1}{2}(\mathbf{Q}^{-\top}\mathbf{C}^{\top}-\mathbf{B}),\\
    \mathbf{S}&=\frac{1}{2}(\mathbf{D}+\mathbf{D}^{\top}), & \mathbf{N}&=\frac{1}{2}(\mathbf{D}-\mathbf{D}^{\top}),
\end{align*}

where it holds by construction that $(\mathbf{J}-\mathbf{R})\mathbf{Q} = \mathbf{A}$, $\mathbf{G}-\mathbf{P} = \mathbf{B}$, $(\mathbf{G}+\mathbf{P})^{\top}\mathbf{Q} = \mathbf{C}$ and $\mathbf{S} + \mathbf{N} = \mathbf{D}$. However, because there exist differently scaled versions of these matrices that describe equivalent dynamics, the constructed matrices are often found to be of significantly different orders of magnitude. Since such differences in the parameterization result in a worse optimization performance \citep{LI_normalization}, a normalization step is added. To be more specific, the Cholesky factorization of $\mathbf{Q}=\mathbf{V}\mathbf{V}^{\top}$ is calculated, giving these final matrices for the linear estimate of the PH system:
\begin{align*}
    \mathbf{J}_{lin}&=\mathbf{V}^{\top}\mathbf{J}\mathbf{V}, & \mathbf{R}_{lin}&=\mathbf{V}^{\top}\mathbf{R}\mathbf{V},\\
    \mathbf{G}_{lin}&=\mathbf{V}^{\top}\mathbf{G}, & \mathbf{P}_{lin}&=\mathbf{V}^{\top}\mathbf{P},\\
    \mathbf{S}_{lin}&=\mathbf{S}, & \mathbf{N}_{lin}&=\mathbf{N}, \\
    \mathbf{Q}_{lin}&= \mathbf{I}_{n_{x}}. & &
\end{align*}

Note that the process described here is for the general case, where a feedthrough term $\mathbf{D}$ is assumed to be present. In the case where such a feedthrough term is not desired, the BLA estimate can enforce the matrix $\mathbf{D}$ to be zero, after which the same approach can be taken, resulting in a linear PH estimate without the $\mathbf{S}_{lin}$ and $\mathbf{N}_{lin}$ terms. Since the original OE-PHNN does not consider a feedthrough term, that scenario will be followed from here on out as well.

\textbf{Direct linear PHNN:}
For the second option, a simpler but more time-consuming alternative can be considered. Instead of applying efficient algorithms for state space estimation, the linear PH estimate is learned by a linear PHNN. Similar to the regular PHNN, a linear PHNN aims to learn the matrices $\jbth, \rbth, \gbth, \pbth$ and the Hamiltonian $H_{\theta}$. In contrast to the nonlinear case however, these matrices do not depend on the current state in the linear case. Therefore, these matrices can be directly parametrized instead of relying on a layered NN.

The downside of this option is that the resulting optimization problem remains nonlinear in the parameters, leaving the possibility that the linear estimate gets stuck in a local optimum. This is less of an issue for the BLA approach, as it is based on a numerically robust subspace identification algorithm.

\begin{figure*}
    \centering
    \includegraphics[scale=0.30]{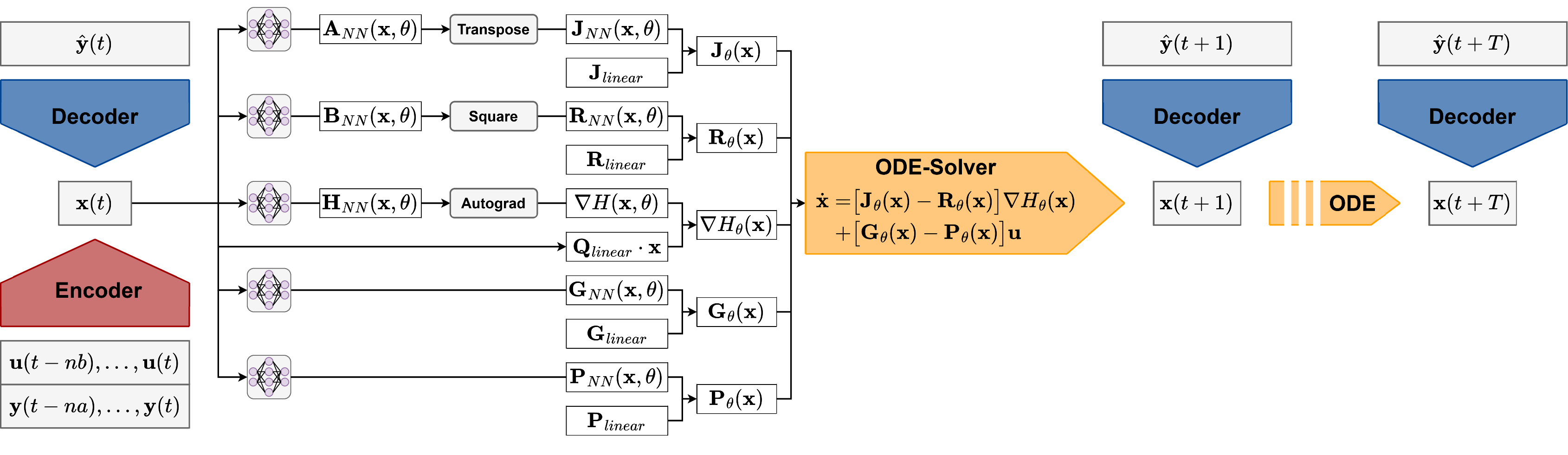}
    \caption{Schematic overview of the model structure including the linear PH estimate.}
    \label{fig:new-structure}
\end{figure*}

\subsection{Initialization of the PHNN}
% How can we use this linear PH for the initialization of the PHNN
% - Where exactly are the linear matrices added?
% - How do we ensure that the initial NN output is 0
In order to use the linear PH estimate as an initialization of the nonlinear PHNN model, a slight modification to the existing structure (see section \ref{sec:PHNN}) must be made. An overview of the adapted model can be seen in Figure \ref{fig:new-structure}.

Instead of calculating the parametrized matrices $\jbth$, $\rbth$, $\gbth$, $\pbth$ and $\hth$ only from the neural networks, the adapted model also includes the linear PH estimates. In practice this simply comes down to summing the terms, i.e.: $\jbth(\xb) = \mathbf{J}_{lin} + \mathbf{J}_{NN}(\xb, \theta)$, $\rbth(\xb) = \mathbf{R}_{lin} + \mathbf{R}_{NN}(\xb, \theta)$, $\gbth(\xb) = \mathbf{G}_{lin} + \mathbf{G}_{NN}(\xb, \theta)$ and $\pbth(\xb) = \mathbf{P}_{lin} + \mathbf{P}_{NN}(\xb, \theta)$.

More attention is required for the gradient of the Hamiltonian however. Since the Hamiltonian itself is represented as a quadratic term for linear PH systems: $H_{lin}(\xb) = \xb^{\top}\mathbf{Q}_{lin}\xb$, its gradient can simply be taken as $\nabla H_{lin}(\xb) = \mathbf{Q}_{lin}\xb$. Therefore, the combined gradient in the adapted model can be formulated as: $\nabla\hth(\xb) = \mathbf{Q}_{lin}\xb + \nabla H_{NN}(\xb, \thb)$.

A second remark needs to be made with regard to the NN parameters $\thb$. If these are still initialized at random, the combination of the linear and NN terms will significantly deviate from the linear estimate, and the advantage of a linear estimate as a starting point will be lost. To prevent this deviation, the NN terms must output zero matrices upon initialization, while still being able to learn the dynamics during training, similar to the approaches presented in \citep{INIT_schoukens1, INIT_schoukens2}. This can be achieved by initializing the weights and biases of the final MLP layer at 0, thus resulting in the terms $\mathbf{J}_{NN}(\xb, \thb)$, etc., being equal to the zero matrix. In this way, the adapted model starts with only the linear estimate, while being able to learn any non-linearities through the NNs during training.

% Ideally would place model structure diagram here
As described, the combination of the linear estimates with the neural network terms for the PH matrices achieves an alternative initialization scheme for the state evolution and decoder terms. Another step towards an improved initialization approach can thus be made by considering the encoder as well, as shown in related work for general subspace encoders \citep{INIT_gerben}.
Ongoing research has been investigating the possibility of initializing the encoder in the model augmentation setting for both linear and nonlinear models \citep{INIT_jan}. This method proposes a direct state estimator from a state-space baseline model, for which an extension to PH systems can be made, but will remain future work for now.
Instead, for this work the encoder is pre-trained on the available dataset from equation \eqref{eq:dataset}, while using only the linear estimate for the state evolution and decoder. Since this encoder pre-training does not require the repeated propagation of the estimated state through time, it can be performed rapidly, only increasing the computational time by a negligible amount.
All in all, combining the data of the nonlinear system with the linear PH estimate in this way yields a suitable initialization for the encoder parameters $\eta$ as well.

\section{Results}
\label{sec:results}
To demonstrate the performance of the proposed identification approach, a numerical implementation is studied. The considered system, consisting of chained mass-spring-dampers, provides a good comparison to existing work on PHNNs and allows for easy experimentation since the data is generated in simulation. Simulations at varying levels of measurement noise are performed to test the method's robustness against noise.
%For the second system, the Cascaded Tanks setup from \citep{RES_cascaded} is considered, to validate the approach on real-world data and allow for comparison to other existing works.

\subsection{Chained Mass-Spring-Dampers}
\begin{figure}
   \centering
   \includegraphics[scale=0.325]{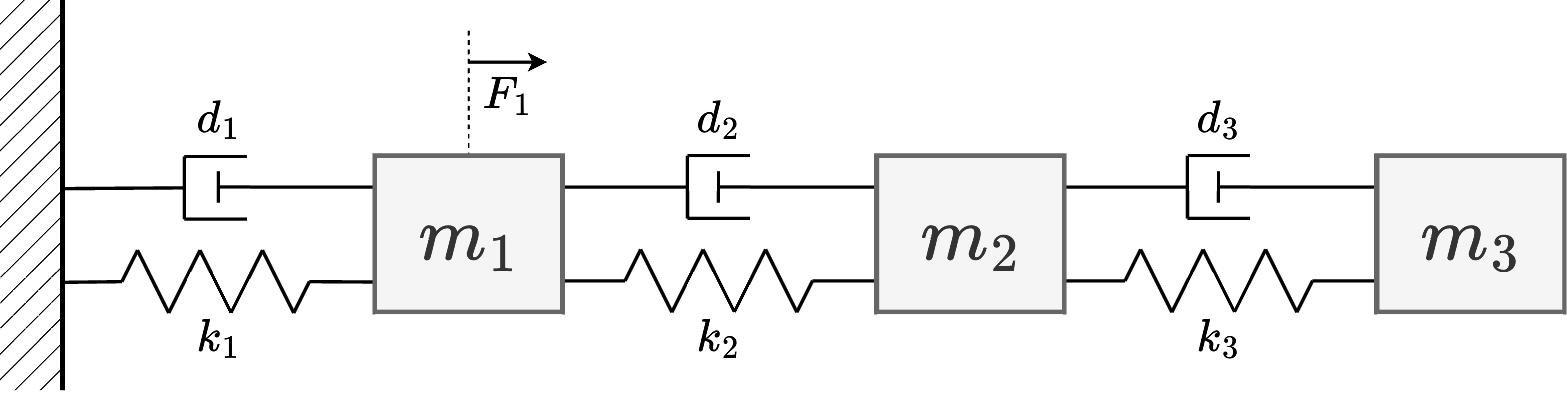}
   \caption{Schematic view of the three coupled mass-spring-dampers. For this system, the input is given as a force $F_{1}$, the states are the displacements, ${q}(t)$, and momenta, ${p}(t)$, of the masses, while the outputs are the velocities of the masses $\dot{{q}}(t)$. Note that the dampers have both a linear and nonlinear component.}
   \label{fig:3-MSD}
\end{figure}

\textbf{Data-generating system:}
% Why do we select chained MSDs?
% - They provide a level training ground, similar system to what the PHNN was developed for.
% - They allow for easy experimentation, since the data is generated in simulation.
Consider a system of three coupled \emph{mass-spring-dampers} (MSDs) with an external force applied to the first mass as illustrated in Fig. \ref{fig:3-MSD}. The system is governed by nonlinear cubic damping dynamics, described by the equation
\begin{equation} \label{eq:3MSD}
    M \ddot{q}(t) + D((\dot{q}(t))^3 + \dot{q}(t)) + Kq(t) = u(t),
\end{equation}
where $M$, $D$, and $K$ are the mass, damping, and spring matrices respectively, while $q(t) \in \mathbb{R}^{3}$ represents the mass displacements. Parameters for all three MSDs are identical and set as $k_{i} = 1$, $m_{i} = 2$, and $d_{i} = 0.5$. A schematic overview of the setup is given in Figure \ref{fig:3-MSD}.

The first mass is driven by a multisine exciting force defined as
\begin{equation}
    \ubt = \sum_{i=1}^{40} \sin(2 \pi i f_{0} t + \phi_{i}),
\end{equation}
with $f_{0} = 0.1$ and random phases $\phi_{i}$ uniformly sampled from $[0, \pi)$. The system is simulated for 100 seconds, with the velocity of each mass sampled at 10Hz, resulting in a dataset of 1000 samples. White noise sampled from a Gaussian distribution is added to the output data to simulate measurement noise. If not mentioned otherwise, the \textit{signal-to-noise ratio} (SNR) is set to 30dB. The resulting dataset consists only of input-output measurements and contains no state observations.\\

% To gauge the ability of the models to deal with noise, datasets are generated at different \textit{signal-to-noise ratio}'s (SNR).

\textbf{Experiments:}
% What kind of results do we want to show?
% - Speed of convergence / consistency across runs
% - Final accuracy (sim/error, all 3 outputs/just 1 output?)
Eight sets of simulation data are generated, each with different initial states and phase realisations of the input signal. Of these experiments, 5 are used for training, 2 for validation, while the final set is kept separate for testing purposes.
To ensure comparability between the original and adapted PHNN approaches, the model structures for the NN terms are kept identical. The encoder is constructed as a residual network with two hidden layers, each containing 64 nodes. The matrices $\jbth$, $\rbth$, $\gbth$, $\pbth$ and $\hth$ are each implemented as a MLP with two hidden layers containing 16 nodes. The default activation function for all networks is the hyperbolic tangent.

All neural networks and parameters were trained using the ADAM optimizer \citep{RES_adam}, with a learning rate of $10^{-3}$, over a fixed duration of $10^{3}$ iterations. Each of those iterations consists of a mini-batch of 64 truncated windows, selected from the training data at random. The encoder is pre-trained for $10^{4}$ iterations with only the linear estimate. RK4 is chosen as the numerical integrator. Overfitting is prevented by selecting the model parameters that achieved the lowest loss on the validation datasets. The results shown in the rest of this section are in relation to the test dataset, which is unseen during the training process. Model performance is quantified by the Normalised Root Mean Square Error (NRMSE),
\begin{equation}
        NRMSE = \frac{\sqrt{\frac{1}{N}\sum_{k=1}^{N} ||y_{k}-\hat {y}_{k}||_{2}^{2}}}{\sigma_y},
\end{equation}
where $\hat{y}_{k}$ is the simulated output, $y_{k}$ the measured output, and $\sigma_{y}$ its standard deviation. \\

\begin{figure}
   \centering
   \includegraphics[width=\linewidth]{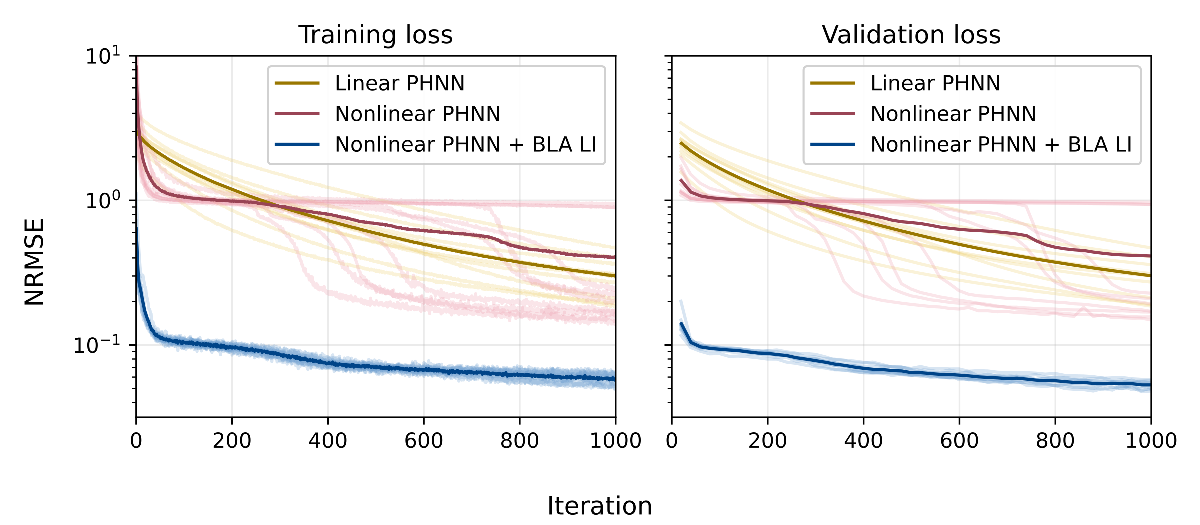}
   \caption{Model convergence during the training process. Yellow shows the linear model, red the nonlinear model and blue the nonlinear model with the linear initialization. The mean for each model class is shown in bold, with individual models opaque.}
   \label{fig:MSD-convergence}
\end{figure}

\textbf{Outcome:}
Figure \ref{fig:MSD-convergence} shows the convergence of the models during the training procedure. It immediately becomes clear from this figure that the linear initialization drastically improves the models' training procedure, as it reduces the deviations between models while also achieving a smaller NRMSE at the end of training. For comparison, while some nonlinear models also achieve low training errors, large differences between the individual models as well as some models getting stuck in local optima, leads to a worse average performance of the nonlinear models. The linear models are more consistent, but do not reach a sufficient accuracy.
%It can be seen that while most nonlinear PHNN models outperform their linear counterparts, some nonlinear models get stuck in a local optimum, drastically reducing the average performance. 

\begin{figure}
   \centering
   \includegraphics[width=\linewidth]{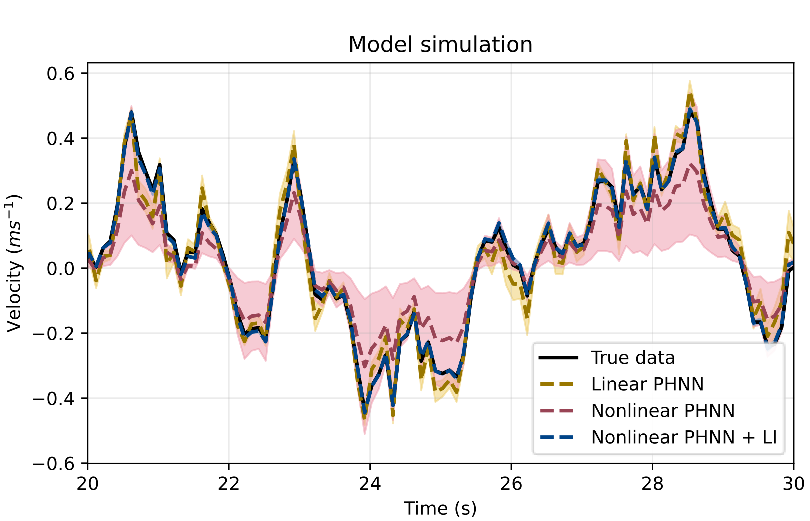}
   \caption{Qualitative simulation of the trained models on a segment of the test dataset. Model class averages are given by the dashed lines, while the shaded areas represent the standard deviations.}
   \label{fig:MSD-quality}
\end{figure}

A similar story can be observed in Figure \ref{fig:MSD-quality}. Here, a small section of the observed output $\dot{q}_{1}$ is compared against each of the methods. Again, the linear initialization drastically reduces the spread in model predictions compared to the nonlinear model. At the same time, while the linear PHNN also shows a small spread between models, it lacks the modelling capacity of the proposed method and does not match the underlying system very well.

\begin{figure}
   \centering
   \includegraphics[width=\linewidth]{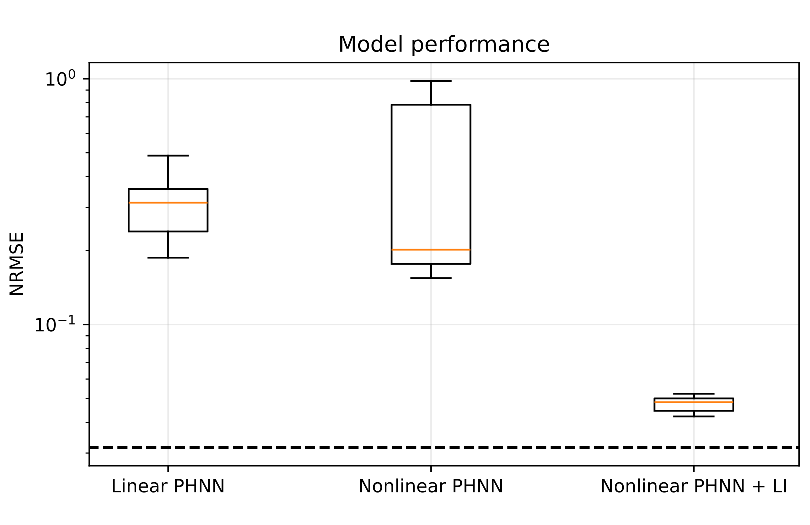}
   \caption{Boxplot showing the spread in model NRMSE's across the considered model classes. The dashed line indicates the noise floor, which is at a SNR of 30dB in this case.}
   \label{fig:MSD-boxplot}
\end{figure}

Additionally, Figure \ref{fig:MSD-boxplot} allows for a quantitative comparison between the model classes. Across 10 trained models for each class, it can be observed that while the linear PHNN has a lower spread between models, it cannot match the accuracy of most nonlinear models. Furthermore, the addition of the linear initialization outperforms the others by reducing both the spread in models as well as the NRMSE on the test set.

Finally, to assess the robustness of the proposed method to measurement noise, a simulation study across various SNR levels is performed. The results of this study are shown in Table \ref{tab:SNR_tests}, where it can be observed that
in the highest noise case (SNR=10dB), the model converges exactly to the noise floor. For the lower noise scenarios however, the simulation errors remain slightly above the noise floor as the optimization has not yet fully converged within the strict $10^{3}$ iteration limit during training. Nevertheless, the spread between models remains relatively small, as the mean and minimum NRMSE's are close for all noise levels.

\begin{table}
    \centering
    \caption{Resulting normalized root mean square errors (NRMSE) for different noise levels in the simulated data.}
    \begin{tabular}{|c|c|c|}
        \hline
        & & \\[-1em]
        SNR & Mean NRMSE & Minimum NRMSE\\
        \hline
        & & \\[-1em]
        10 dB & $ 3.09 \cdot 10^{-1}$ & $ 3.08 \cdot 10^{-1}$ \\
        20 dB & $ 1.13 \cdot 10^{-1}$ & $ 1.10 \cdot 10^{-1}$ \\
        30 dB & $ 4.22 \cdot 10^{-2}$ & $ 3.98 \cdot 10^{-2}$ \\
        40 dB & $ 2.03 \cdot 10^{-2}$ & $ 1.90 \cdot 10^{-2}$ \\
        \hline
    \end{tabular}
    \label{tab:SNR_tests}
\end{table}

\section{Conclusion}
\label{sec:conclusion}
This paper presents a novel initialization strategy for output-error port-Hamiltonian neural networks. It considers both a direct and an indirect method to acquire a linear PH estimate of a system, which can then be used as a starting point for the neural networks to learn the system nonlinearities. The performance of this initialization is then compared against the original method in numerical experiments. For the first case study, a simulated mass-spring-damper dataset is considered, for which the proposed method significantly decreases both the spread between models as well as the final NRMSE.
Estimates reliably converge towards the noise floor across different signal-to-noise ratio's, indicating that the proposed method effectively improves the training procedure of OE-PHNNs.

\bibliography{ifacconf}

\end{document}